\documentclass[runningheads]{svmult}

\usepackage{makeidx}   % allows index generation
\usepackage{graphicx}  % standard LaTeX graphics tool
\usepackage{subeqnar}  % subnumbers individual equations
\usepackage{multicol}  % used for the two-column index
\usepackage{physprbb}  % modified textarea for proceedings,
\makeindex             % used for the subject index
\newcommand{\greeksym}[1]{{\usefont{U}{psy}{m}{n}#1}}
\newcommand{\umu}{\mbox{\greeksym{m}}}

\begin{document}
\title*{Linear Spectropolarimetry of Young and\protect\newline Other
Emission Line Stars}
\toctitle{Linear Spectropolarimetry of Young and
\protect\newline Other Emission Line Stars}
% allows explicit linebreak for the table of content
%
%
\titlerunning{Linear Spectropolarimetry of YSOs}
% allows abbreviation of title, if the full title is too long
% to fit in the running head
%
\author{Janet E. Drew\inst{1}
\and Jorick S. Vink\inst{1}
\and Tim J. Harries\inst{2}
\and Ryuichi Kurosawa\inst{2}
\and Ren\'e D. Oudmaijer\inst{3}}

\authorrunning{J. E. Drew et al.}
% if there are more than two authors,
% please abbreviate author list for running head
%
%
\institute{Imperial College London, Physics Department, Exhibition Road,\\
           London SW7 2AZ, UK
\and University of Exeter, School of Physics, Stocker Road, Exeter EX4 4QL, UK
\and University of Leeds, School of Physics \& Astronomy, EC Stoner 
Building,\\
           Leeds LS2 9JT, UK}

\maketitle              % typesets the title of the contribution

%\begin{abstract}
%Text here....
%\end{abstract}

\section{Introduction}
The aim of this article is to demonstrate the useful role that can be
played by spectropolarimetric observations of young and evolved emission
line stars that analyse the linearly polarized component in their
spectra.  At the time of writing, this demonstration has to be made on
the basis of optical data since there is no common-user infrared facility,
in operation, that offers the desired combination of spectral resolution 
and sensitivity.  If the new ESO instrument, CRIRES, can be characterised
to sufficient precision, it may become the first capable of performing
such science.

The case for high signal-to-noise (S/N) spectropolarimetry on the largest
available telescopes has been made before at an earlier ESO conference
(\cite{Schmid}, \cite{Donati}).  Because the polarised fraction of the 
light received from astrophysical sources is often only of the order of 1\% or 
so, it is clear that observations seeking to separate out and characterise this
component suffer at least a 5-magnitude disadvantage relative to total light
spectroscopy: expressed in S/N terms, one can only begin to achieve anything 
at S/N approaching 1~000.  It was shown by both Schmid et al~\cite{Schmid} and 
Donati et al~\cite{Donati} that some of the most exciting results, 
particularly from the analysis of circular-polarised light, will only come 
as the achievable S/N rises to 100~000.

Here we focus on what can be learned from linear spectropolarimetry alone 
at reasonably high spectral resolution (R of order 20000 is already valuable) 
and at $10^3 < $S/N$ < 10^4$.  And we remind that the near infrared 
(1$\umu$m--2$\umu$m) has the potential to out-perform the optical as a domain 
to work in because of the greatly reduced interstellar obscuration at these 
wavelengths.

\section{The astrophysical rewards of higher spectral resolution linear 
spectropolarimetry}

Is the 'waste' of observing for so long, that the information content of a mere
percent or so of the received light can be dragged out of it, worth it?  It is,
and one can understand how by the following argument.  

First, the spatial scales on which stars and their immediate environments 
(accretion flows, winds, magnetic structures, disks...) are all too often too 
compact to be directly imaged.  This is a state of affairs that will persist 
even beyond the commissioning of 100~m class telescopes or advanced 
interferometers aiming to get down to the milli-arcsec scale at OIR 
wavelengths. Second, spectroscopy aiming to deduce unique geometric and 
kinematic information from one or more well-resolved spectral line profiles 
will often run into the sand of information loss due to the essential 
convolution that underlies spectral line formation -- the more complex the 
line-forming environment, the worse it gets.  

In such circumstances, the role of linear spectropolarimetry is its access
to a different dimension of information: in particular, linear polarized
light is in most cases scattered light, whilst unpolarized light is directly
sourced light (e.g. coming directly from the star, or from the 
$\tau_{\lambda} \simeq 1$ surface within a line-forming diffuse gas).  It
has been recognised for over 40 years that the analysis of linearly-polarised
light can provide vital clues to otherwise unresolved geometries.  A telling
success of the technique was in its role in underpinning the `unification'
model of active galactic nuclei~\cite{A93}.

Here, the discussion centres on applications to young and other emission
line stars.  The great early success of linear spectropolarimetry in this
context was in motivating the now widely-accepted circumstellar disk model 
for the (still perplexing) classical Be stars~\cite{PR03}.  Our own optical 
observations of young stellar objects (YSOs), summarised below, were inspired 
by this example.  An important point of contrast between stellar and AGN 
applications of linear spectropolarimetry is that the former place much higher 
demands on spectral resolution: emission line widths are on the order of 
100 km~s$^{-1}$, and the goal is to trace polarization changes \emph{within} 
them.

\section{The current optical state of the art}

\begin{figure}
\begin{center}
\includegraphics[width=.72\textwidth]{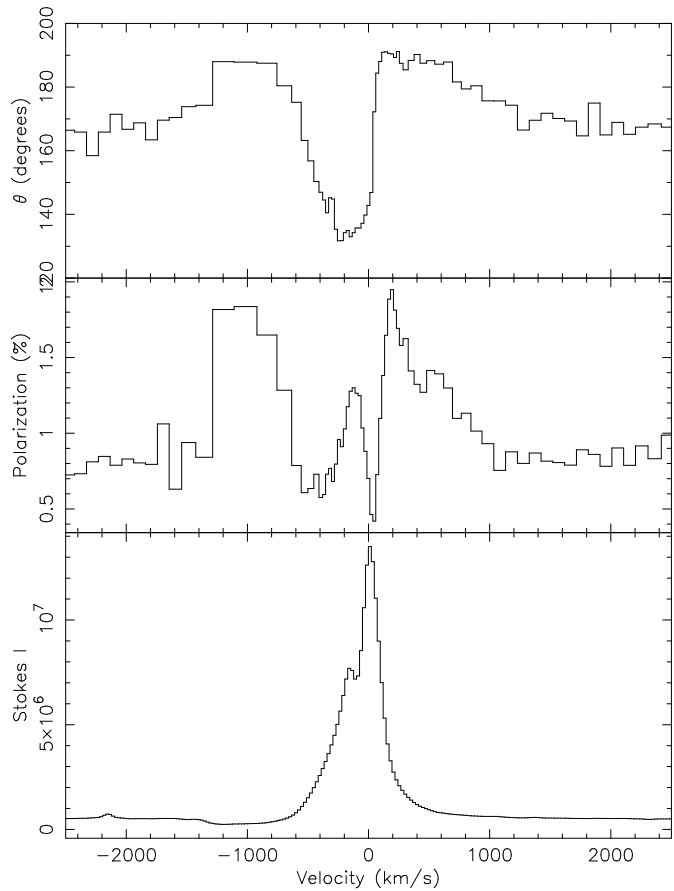}
\end{center}
\caption[]{H$\alpha$ polarisation of the B[e] star HD 87643 (Oudmaijer
et al~\cite{OPDW}).  This bright B[e] star with its broad H$\alpha$ 
profile amounts one of a limited number of stars that fall well within the
grasp of the 4-metre class telescopes. S/N would benefit from some 
improvement, but the spectral resolution achieved ($\sim$5000) is 
adequate.  The binning is adaptively set to yield 0.1\% errors.}
\label{eps1}
\end{figure}

The beginnings of optical high resolution spectropolarimetry, aiming to reveal
linear polarization changes within emission line profiles date back to the
1970s (\cite{PM77}, \cite{MC78}, \cite{MCFS}). This was very hard work, 
verging on the heroic.  For example, the H$\beta$ observations of 4 bright 
classical Be stars obtained by McLean et al (1979)~\cite{MCFS} were the fruit 
of 3 nights on a 1.54-m telescope: sampled on a bin-size of 0.45~\AA\, the
data gave a good, if a little noisy, illustration of a phenomenon common to 
this object class that is sometimes referred to as `depolarization' (see
below).  Things have got better.

Jumping two decades, we present an example (Figure~\ref{eps1}) 
illustrating the gain in information to be had from well-resolved 
high-S/N linear spectropolarimetry.  This is an AAT observation of 
the B[e] star, HD~87643 at H$\alpha$ (from \cite{OPDW}). It has come out so 
well both because the target star is optically bright ($V = 9$) and the 
H$\alpha$ line profile is relatively broad, with FWHM exceeding 500 
km~s$^{-1}$: a resolving power of just 5000 is sufficient for a good result.  
In this case the polarimetric changes across the line are quite complex.
The position angle rotation, in particular, leads to an interpretation in 
terms of scattering of a compact emission line source by a rotating medium 
such as a circumstellar disk (cf Wood et al~\cite{WBF}).  In addition HD~87643 
affords a good example of what we refer to as the `McLean effect' (described 
and explained by McLean (1979)~\cite{M79}), wherein the P~Cygni absorption 
seen in Stokes I is accompanied by a clear change in the percentage 
polarization: this can be seen as the line absorption having preferentially 
removed direct, unscattered starlight from the beam.  These data are 
certainly of a quality that justifies comparison with state-of-the-art 
radiation transfer modelling, but the number 
of optically-accessible objects that can be observed to a similar standard
is limited, and of necessity smaller than the number accessible at 
1$\umu$m--2$\umu$m.

\section{A particular campaign: probing the circumstellar environments of YSOs}

   In the past few years we have begun to explore the circumstellar
environments (CSM) of YSOs via a sequence of H$\alpha$ linear 
spectropolarimetric observing programmes using 4-metre class telescopes 
(the AAT and WHT).

   The first objective was the relatively limited one of seeing how far 
the brightest of the optically-accessible YSOs, selected from among the known 
Herbig~Be stars, followed the same polarimetric pattern of behaviour as the 
classical Be stars~\cite{PM76}.  We saw this as a simple, if 
statistically-based, way of providing further evidence for or against the 
concept that even Herbig~Be stars remain embedded in circumstellar disks -- 
like classical Be stars.  The background to this is that doubt has been 
expressed that such objects can still be in possession of disks (see e.g. the 
review by Waters \& Waelkens~\cite{WW98}, and the results of NIR 
\begin{figure}
\begin{center}
\includegraphics[width=.6\textwidth]{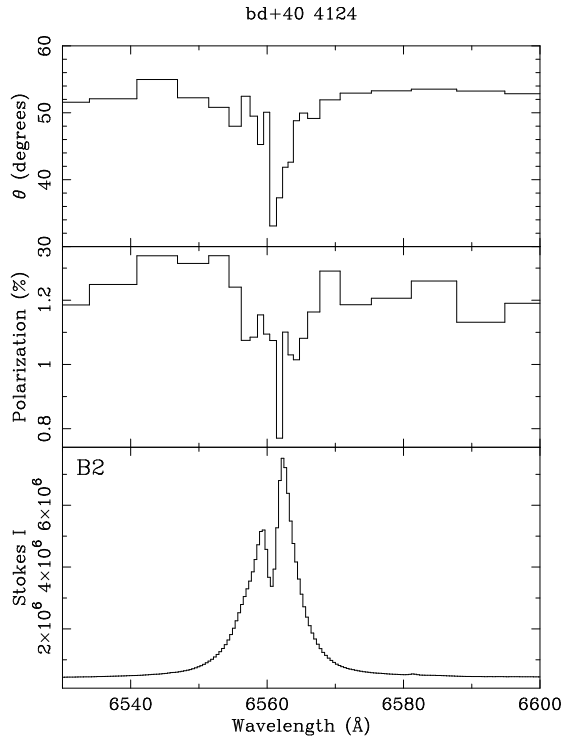}
\end{center}
\caption[]{Linear polarisation at H$\alpha$ of the Herbig Be star, 
BD$+$40$^{\rm o}$~4124: an example of the `depolarization' effect where
the line emission serves mainly to `dilute' intrinsic continuum 
polarization.  The change across the line profile is smooth in one or 
both of the percentage and the position angle, crudely mimicking the
shape of the Stokes I profile. The binning is set to yield 0.05\% 
errors.}
\label{eps6}
\end{figure}
\begin{figure}
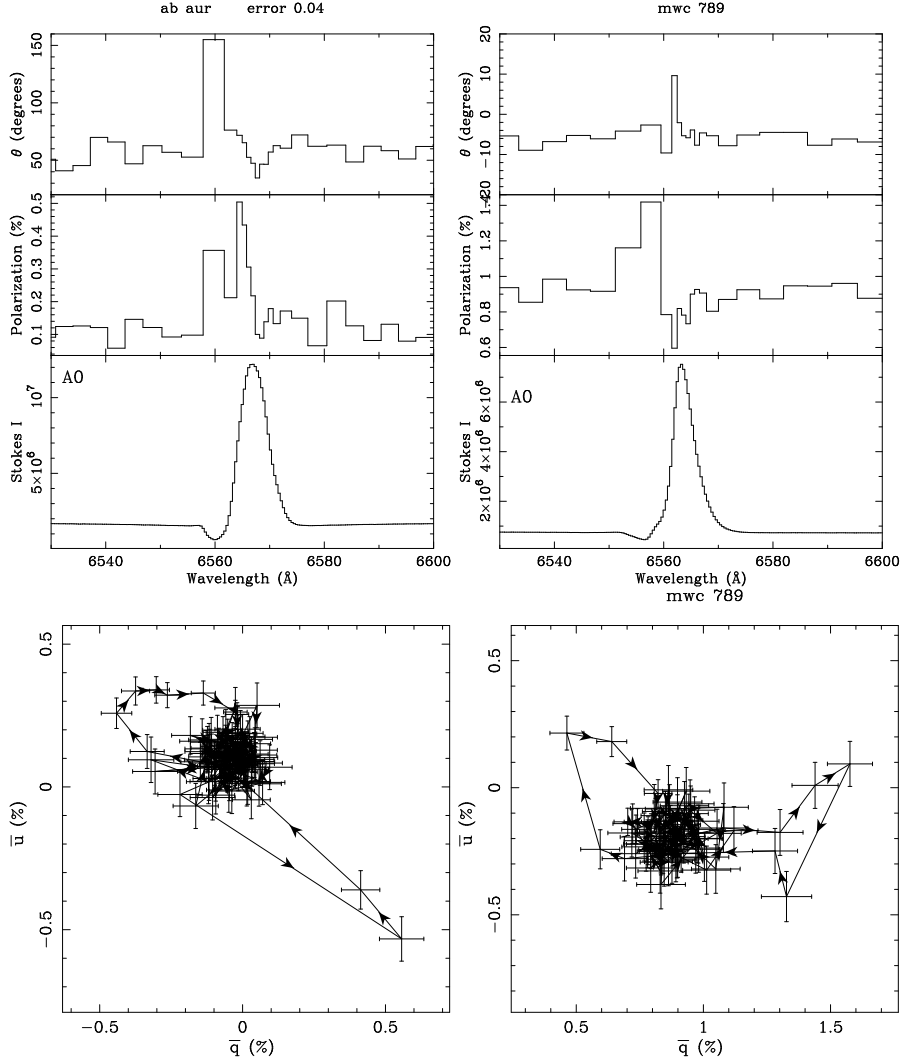

\begin{center}
\includegraphics[width=.48\textwidth]{drewF3tl.epsi}
\includegraphics[width=.48\textwidth]{drewF3tr.epsi}
\includegraphics[width=.48\textwidth]{drewF3bl.epsi}
\includegraphics[width=.48\textwidth]{drewF3br.epsi}
\end{center}
\caption[]{H$\alpha$ polarisation of the Herbig Ae stars AB Aur and 
MWC 789.  Both stars have been assigned A0 spectral types and present
moderately P~Cygni H$\alpha$ Stokes I line profiles. Further
qualitative similarities in the polarization domain are the `McLean effect' 
in the percentage linked to the Stokes I blueshifted absorption, and the 
position angle rotations linked to the redshifted emission.  The additional
polarization data quickly identify different on-sky angles and, in
the QU plane, there are quantitative differences that would warrant
investigation via numerical modelling. The error per bin is: top left
0.04\%; top right 0.05\%; bottom left 0.08\%, bottom right 0.1\%.}
\label{eps2}
\end{figure}
interferometry presented by Millan-Gabet et al~\cite{MG01}).  The upshot was
that we found `depolarization' across H$\alpha$ in 7 out of 12 Herbig Be 
stars observed (\cite{V2002}).  Down to the same sensitivity limit, this is 
a rate of incidence matching that associated with classical Be stars.
The term `depolarization' (illustrated in figure~\ref{eps6}) describes a 
smooth polarization change across the emission line resulting from the 
combination of (i) intrinsic linear polarization of the stellar continuum 
due to 
scattering in a non-spherical CSM, with (ii) no polarization of the line 
emission which itself forms in the larger CSM, {\em other} than by the 
ISM intervening between YSO and observer that polarizes line and adjacent 
continuum equally.  Our result thus points toward broad comparability between 
the two object groups and strengthens the case for what are most likely
electron-scattering disks around Herbig Be stars.

   It was natural to extend this programme down to less massive and luminous
Herbig Ae stars to see if the linear polarization signature, attributable to 
the CSM, remains the same or changes.  From the purely observational point of
view this extension turned out to be very rewarding in that the proportion
of stars showing a clear polarization change across the H$\alpha$ line 
increased relative to the Herbig Be star: 9 out of the 11 Herbig Ae stars
observed showed changes, and in every case there was a distinctive position
angle flip (cf HD~87643) not generally seen in the Herbig Be stars (see
figure~\ref{eps2} and also \cite{V2002}).  This behaviour implies intrinsic 
polarization of some fraction of the H$\alpha$ emission itself -- most likely 
due to H$\alpha$ emission in the immediate vicinity of the star (magnetospheric
accretion?) scattering off a larger rotating CSM (a dust-bearing 
disk?).  The very fact that this
\begin{figure}
\begin{center}
\includegraphics[width=.7\textwidth]{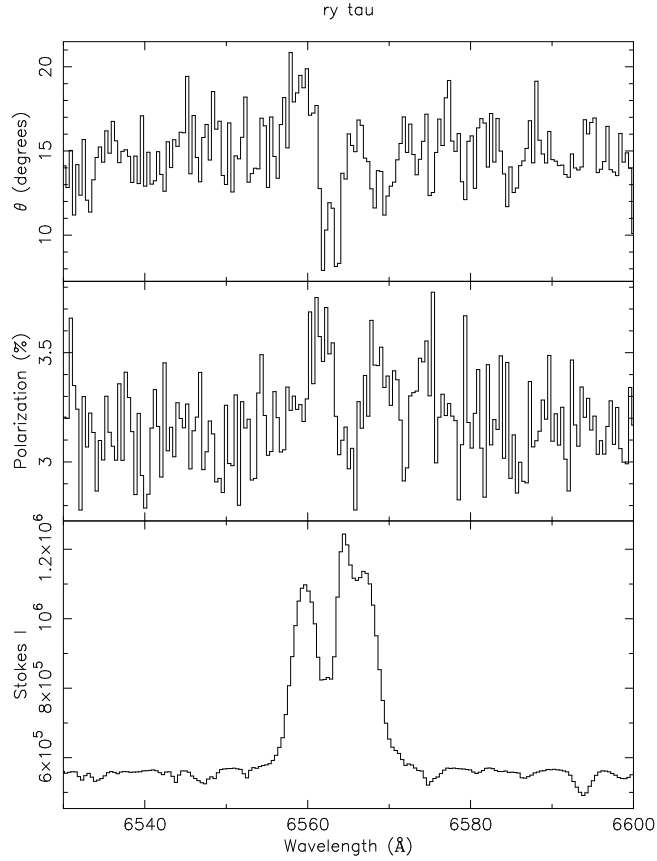}
\end{center}
\caption[]{H$\alpha$ polarisation of the classical T Tauri star (F8) star
RY Tau. In this case the data have not been rebinned to a fixed error level
but are shown at the detector binning. See \cite{V2003} for a discussion
and interpretation of these data -- obtained, like most of the Herbig
star data, using ISIS$+$polarimetry module on the WHT. }
\label{eps3}
\end{figure}
behaviour is so common indicates that, despite a widely held and discouraging 
impression of great optical depth in the H$\alpha$ line in Herbig~Ae and 
T~Tau spectra, we nevertheless typically gain a partially unhindered view in 
this line of the very compact scale of one-to-a-few stellar radii.  The view 
via the less-opaque near-infrared H~{\sc i} lines can only be even better.

   At optical (H$\alpha$) wavelengths the T~Tau stars, at still lower mass
and luminosity, are much more challenging targets for observation.  
Figure~\ref{eps3} shows H$\alpha$  in RY~Tau, our best example from this 
object class to date (\cite{V2003}).  To achieve S/N of at least 1000 per 
sample, the observational requirement is 2 million photons: in RY Tau 
($V \simeq 10$) we managed half this per resolution element in the 
continuum at R $\simeq$ 9000 in one hour on the WHT.  Unfortunately nearly 
all T~Tau stars are a couple of magnitudes or more fainter than this!  We are 
working to gather what data we can from a handful of T~Tau stars, using 
4-metre telescopes -- but, in truth, the need to switch to larger 
telescopes is reached with this object class.  Furthermore, many T~Tau stars 
are significantly reddened and hence much brighter and better tackled at 
infrared wavelengths.

\begin{figure}
\begin{center}
\includegraphics[width=.6\textwidth]{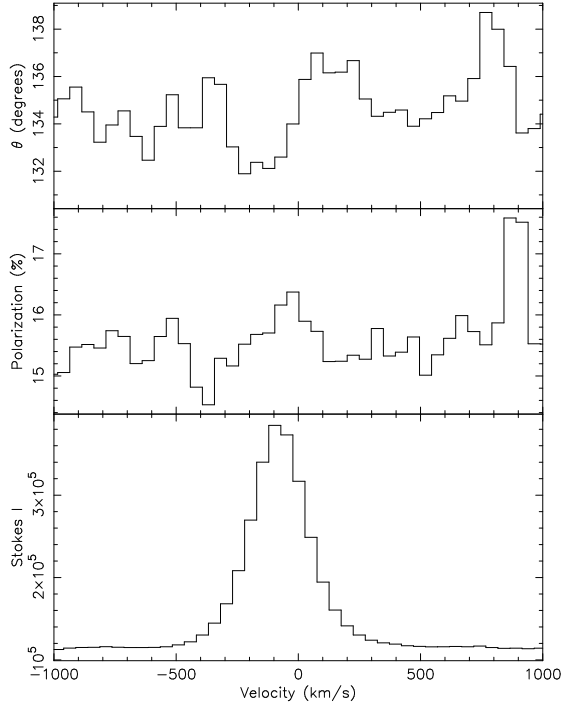}
\end{center}
\caption[]{Linear polarization data at Pa$\beta$ for the embedded massive 
Young Stellar Object GL 490 (otherwise unpublished).  Note the tantalising
glimpse of a position-angle rotation across the emission line -- accompanied
elsewhere in the spectrum by less welcome (inexplicable) excursions. The
error per bin is 0.1\%.}
\label{eps4}
\end{figure}

   Finally -- what of massive YSOs and other deeply-embedded populations
of young objects?  We have attempted one or two of these at near-infrared 
wavelengths using a 4-metre class telescope (CGS4 on UKIRT). Shown as 
figure~\ref{eps4} is an observation at Pa$\beta$ of the embedded massive
YSO, GL~490: it is noisy, it would benefit from higher spectral resolution
and yet offers promise by way of what might be a position angle flip
across the H~{\sc i} line.  The way forward is clear -- higher resolution
and more photons from a suitable instrument on the VLT or other large
telescope.

\begin{figure}
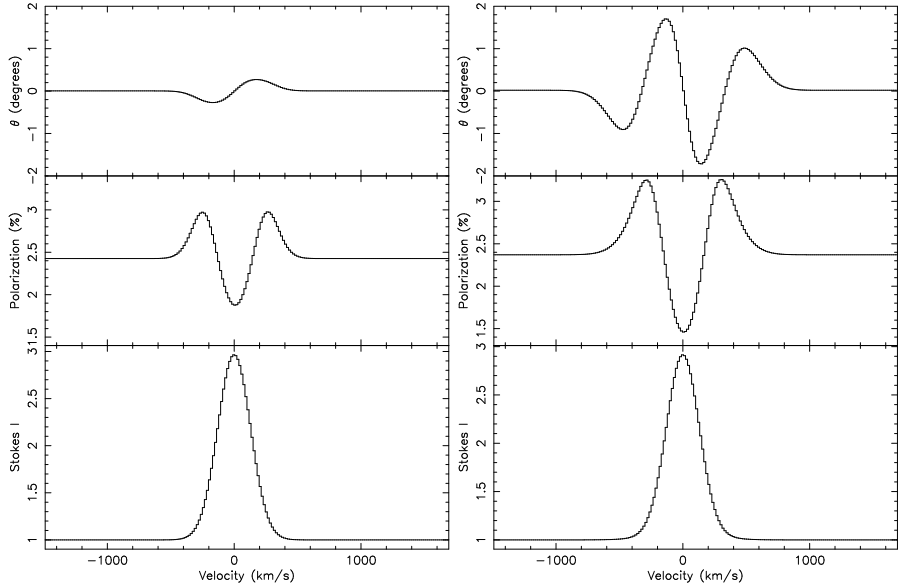

\begin{center}
\includegraphics[width=.48\textwidth]{drewF6l.epsi}
\includegraphics[width=.48\textwidth]{drewF6r.epsi}
\end{center}
\caption[]{Monte Carlo predictions of the H$\alpha$ line polarisation 
for the cases of a finite-sized line-emitting star embedded in a scattering
disk with (left) and without an inner hole (right). The disk is inclined at
45$^{\rm o}$ to the line of sight and, in the left panel, has an inner hole
of radius 5 times the stellar radius.}
\label{eps5}
\end{figure}

\section{A new age of theoretical modelling}

    Modelling of the circumstellar environments of YSOs and related objects
begins from the general concept that the linearly polarized
light component in a spectrum results from scattering within a non-sperical
medium.  The scatterers may be circumstellar electrons (as is likely for 
more massive, luminous young objects) or dust (both high- and low-mass YSOs). 
Although simple analytical treatments of circumstellar scattering have proved 
to give useful insights into circumstellar geometries (e.g. \cite{BM77},
\cite{WBF}) more sophisticated modelling, particularly of 
optically-thick structures and non-idealised geometries, requires a numerical 
approach. The Monte-Carlo method combined with the high power of today's 
desk-top computing provides a straightforward and viable techique for 
modelling radiative transfer even in multiple-scattering, three-dimensional 
environments.

We are continuing to develop the radiative-transfer model {\sc torus}
(Harries 2000~\cite{H00}) in order to do two things: (i) extend the simplified 
analytic treatments like those of Wood et~al \cite{WBF} into a geometrically
more flexible numerical domain to verify and enrich our basic insights, 
and (ii) produce realistic models of line and continuum formation in YSOs for 
more exacting comparison with observations. In both approaches, the 
radiative-transfer is followed in a three-dimensional grid, which may be 
described by spherical polar or cartesian coordinates, or via a cubic 
adaptive mesh.

Figure~\ref{eps5} shows an intriguing result that is emerging from even
the more simplified of the two approaches, that merely replaces analytic
ideal structures with real 3-D geometry (with no physics of the
microstate).  We are finding that there is a, hitherto unsuspected, 
qualitative difference between scattering of the light from a line- and 
continuum-emitting star by, on the one hand, a rotating circumstellar disk 
reaching into the stellar surface and, on the other, a rotating disk with a 
significant central hole.  The single position-angle flip, with two turning
points, seen in the second case is as predicted by analytic models -- 
but the doubling to four turning points associated 
with the undisrupted disk reaching into the surface is the surprise. 
In-depth numerical exploration of this result shows it is due, indeed, to 
the geometrically correct treatment of the finite sized star interacting with 
the disk's rotational velocity field (which re-sorts the scattered line 
emission).

\begin{figure}
\begin{center}
\includegraphics[width=\textwidth]{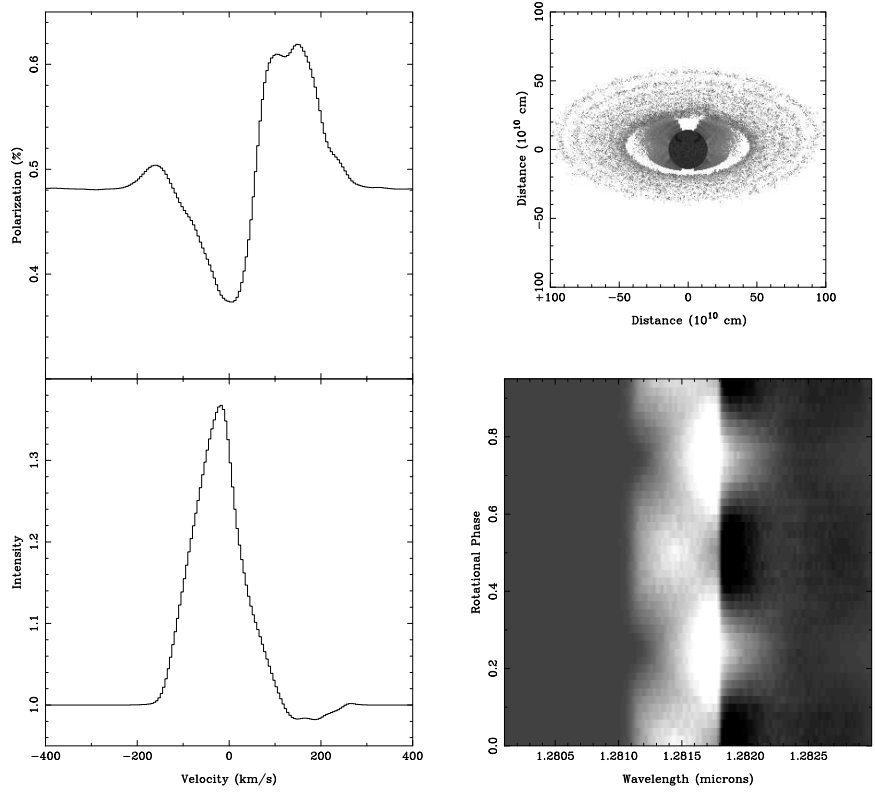}
\end{center}
\caption[]{A Pa$\beta$ spectropolarimetric model of magnetospheric accretion in
a typical CTTS. The left-hand panel shows a typical Pa$\beta$ polarization 
spectrum. The top-right panel shows a logarithmic greyscale image of the 
Pa$\beta$ intensity (white is dark, grey is bright): note the two accretion 
curtains and the scattered light component from the inner-edge of the disc. 
The dynamic time-series spectrum is also plotted (lower right, with emission
shown as white, and absorption as dark). }
\label{tims}
\end{figure}

With a view to direct comparison of models with data, Harries, Kurosawa and
Symington are applying the code to the computation of emission line profiles 
due to magnetospheric accretion in T~Tauri stars.  This fuller treatment
requires that the equation of statistical equilibrium for hydrogen is solved 
for each cell in the grid, while the line transfer is computed under the 
Sobolev approximation.  The dust radiative-equilibrium may be simultaneously 
solved using the MC method of Lucy~\cite{L99}.  By introducing 
azimuthal structure into the geometry we are able to compute synthetic 
time-series spectra. There is no significant linear polarization from the 
magnetosphere alone, but if we 
introduce a simple circumstellar disc we find signficant levels of line and 
continuum polarization. The structure of the polarization through the line 
provides a potentially important lever on the circumstellar structure, since 
the scattered component arises from the inner-edge of the disc which 'sees' the
magnetosphere from a distribution of viewing angles that are different
to that of the observer (integrating the direct light). In order to give a 
flavour of the sophistication of the models in figure~\ref{tims} we present 
a sample model of a structure magnetosphere, with two diametrically opposed
accretion 'curtains' of 120 degrees extent.

The combination of models of this type, employing self-consistent physics 
with high-quality spectropolarimetry, will provide tight constraints on
the structure and dynamics of the accretion flow and/or CSM in both 
high- and low-mass YSOs. And of course the approaches sketched here can be
adapted to the analysis of other object classes. 

\section{Final words}

This discussion of the application of high resolution linear 
spectropolarimetry to emission line stars has emphasised studies of young
objects.  Other important applications would be to studies of intermediate
mass (post-AGB) stars and high mass evolved stars.  Regarding the latter
there are pressing issues concerning the role of angular momentum and how
this expresses itself in e.g. B[e] stars (a more extreme and evolved type
relative to the classical Be stars).  In all these applications the
near-infrared is clearly a welcoming wavelength range. Optical work to date 
has been limited to resolving powers at or below $\sim$10~000.  We can see 
now that a factor of 2 or 3 improvement is desirable (e.g. figure~\ref{eps2}), 
and also some improvement in S/N beyond 1000 -- which has been attained at 
times!  This point has been reached at a time when theory, exploiting flexible 
Monte Carlo methods, is fast becoming a powerful tool.  In short we have the 
complex phenomena, and the rise of the modelling capability to match -- good 
data are the missing link.

%INDEX%%%%%%%%%%%%%%%%%%%%%%%%%%%%%%%%%%%%%%%%%%%%%%%%%%%%%%%%%%%%%%%
% Please check with the editor of your book whether he plans to
% include a "mutual" subject index - if so, please code your entries
% in the standard syntax. For your own purposes you may print your
% "personal" index by using the following commands:
%
%\clearpage
%\addcontentsline{toc}{section}{Index}
%\flushbottom
%\printindex
%%%%%%%%%%%%%%%%%%%%%%%%%%%%%%%%%%%%%%%%%%%%%%%%%%%%%%%%%%%%%%%%%%%%%

\end{document}